\documentclass[twocolumn,showpacs,preprintnumbers,amsmath,amssymb,superscriptaddress]{revtex4}

\usepackage{graphicx}
\usepackage{dcolumn}
\usepackage{bm}
\usepackage{subfigure}

\begin{document}


\title{Tunneling-induced coherent electron population transfer in an asymmetric quantum well}

\author{Ni Cui}
\affiliation{State Key Laboratory of High Field Laser Physics,
Shanghai Institute of Optics and Fine Mechanics, Chinese Academy of
Sciences, Shanghai 201800, China}%
\affiliation{Graduate University of Chinese Academy of Sciences,
Beijing 100049, China}
\author{Yueping Niu}\thanks{Corresponding author.
E-mail: niuyp@mail.siom.ac.cn}%
\affiliation{State Key Laboratory of High Field Laser Physics,
Shanghai Institute of Optics and Fine Mechanics, Chinese Academy of
Sciences, Shanghai 201800, China}%
\author{Shangqing Gong}\thanks{Corresponding author.
E-mail: sqgong@mail.siom.ac.cn}%
\affiliation{State Key Laboratory of
High Field Laser Physics, Shanghai Institute of Optics and Fine
Mechanics, Chinese Academy of
Sciences, Shanghai 201800, China}%

\date{\today}

\begin{abstract}
We propose an asymmetric double quantum well structure with a common
continuum and investigate the effect of resonant tunneling on the
control of coherent electron population transfer between the two
quantum wells. By numerically solving the motion equations of
element moments, the almost complete electron population transfer
from initial subband to the target subband could be realized due to
the constructive interference via flexibly adjusting the structure
parameters.
\end{abstract}

\maketitle

\section{introduction}

Coherent population transfer among discrete quantum states in atoms
and molecules has attracted tremendous attentions\cite{CPT, TPT}
over the past decades, due to its potential application to atomic
optics\cite{AO}, preparation of entanglement\cite{Entangle} and
quantum computation\cite{QC}. The most robust technique for
achieving efficient population transfer is stimulated Raman
adiabatic passage (STIRAP)\cite{CPT, TPT, PS}, by which perfect
population transfer has been studied in a three-level $\Lambda$
system particularly. Moreover, the case when the upper isolated
discrete level is replaced with a continuum was suggested initially
by Carroll and Hioe\cite{Carroll-Hioe}. In their model, the
continuum could serve as an intermediary for population transfer
between two discrete states in an atom or a molecule by STIRAP.
However, complete population transfer is unrealistic, significant
partial transfer may still be feasible\cite{Chirp,tripod}.

Recently, there is great interest in extending these studies to
semiconductor nanostructures\cite{UAPT, QD, QD1, Biexciton} for the
possible implementation of optoelectronic devices. In the conduction
band of semiconductor quantum structure, the confined electron gas
exhibits atomiclike behaviors. The intersubband energies and
electron wave function symmetries could be engineered, with a
flexibility unknown in atomic systems, to suit particular
requirements. Therefore, several quantum optical coherence and
interference effects have been investigated within intersubband
transitions of semiconductor quantum wells(QWs), such as
electromagnetically induced transparency(EIT)\cite{EIT},
tunneling-induced transparency\cite{TIT}, enhancement of Kerr
nonlinearity\cite{Kerr, XPM}, ultrafast all-optical
switching\cite{UOS}, four-wave mixing\cite{FWM}, Rabi
osillations\cite{RO}, and coherent population
trapping(CPT)\cite{PT}.

In the present paper, we design a n-doped asymmetric AlGaAs/GaAs
double QW with a continuum(Fig.~\ref{fig:1}), which has four
conduction subbands with a closely separated excited doublet. It is
well known that, for an excited-doublet four-level atom system, the
quantum interference effect can lead to depression or even
cancellation of spontaneous emission from the excited doublet to the
lower levels\cite{Scully, Knight, Li}. Considering above, some of
us\cite{Gong} demonstrated that a complete population transfer from
an initial state to another target state and an arbitrary
superposition of atomic states could be realized with the STIRAP
technique. However, the spontaneous decay is intrinsically
incoherent in nature and detrimental to coherent population
transfer, Zhu \emph{et al.}\cite{Zhu} investigated that the
remarkable enhancement or suppression of population transfer can be
realized through the spontaneous emission constructive or
destructive quantum interference even with a finite pulse area. In
contrast to the excited-doublet four-level atom, in our structure,
the quantum tunneling to a continuum from the excited doublet could
give rise to the Fano-type interference\cite{Fano}, which has been
observed experimentally in semiconductor intersubband
transitions\cite{Fano-Ex} and the sign of quantum interference
(constructive or destructive) in optical absorption can be reversed
by varying the direction of tunneling from the excited doublet to a
common continuum\cite{Fano-sign}. Therefore, in the present paper,
we will analyze the controllability of the coherent electron
population transfer in the four-subband double QW structure via
tuning the sign of quantum interference due to resonant tunneling,
the splitting of the excited doublet (the coupling strength of the
tunneling), and the Fano-type interference. By numerically solving
the motion equations of element moments, the almost complete
electron population transfer between QWs could be realized due to
the constructive interference by suitably adjusting the structure
parameters.

\section{The model and basic equations}

\begin{figure}[htbp]
 \includegraphics[width=7cm, height=5cm]{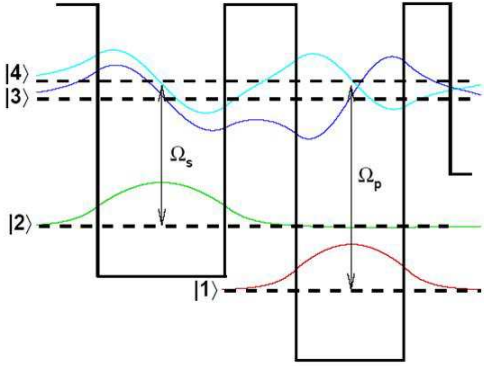}
   \caption{\label{fig:1}Conduction subbands of the asymmetric double QW.
   It consists of two wells and a collector region
   separated by thin tunneling barriers. The dash lines denote
   the energy subbands and the solid curves represent the
   corresponding wave functions.}
\end{figure}

The \emph{n}-doped asymmetric GaAs/AlGaAs double QW under
consideration is shown in Fig.~\ref{fig:1}. One ${\rm Al}_{0.07}{\rm
Ga}_{0.93} {\rm As}$ QW with thickness of $8.3\ \rm nm$ is separated
from a $6.9\ \rm nm$ GaAs QW by a $4.8\ \rm nm$ ${\rm Al}_{0.32}{\rm
Ga}_{0.68}{\rm As}$ potential barrier. On the right side of right
well is a $3.4\ \rm nm$ thin ${\rm Al}_{0.32}{\rm Ga}_{0.68}{\rm
As}$ barrier, which is followed by a thick ${\rm Al}_{0.16}{\rm
Ga}_{0.84}{\rm As}$ layer. In this structure, the energies of the
ground subbands $|1\rangle$ and $|2\rangle$ for the two wells are
obtained as $51.53\ \rm meV$ and $97.78\ \rm meV$, respectively. Two
closely spaced delocalized upper levels $|3\rangle$ and $|4\rangle$,
with the energies $191.30\ \rm meV$ and $203.06\ \rm meV$
respectively, are created by mixing the first excited subband of the
shallow well $|se\rangle$ and the first excited subband of the right
deep well $|de\rangle$ by tunneling. The dash lines denote the
energy subbands and the solid curves represent the corresponding
wave functions. The subband $|1\rangle$ and subbands $|3\rangle$ and
$|4\rangle$ are coupled by the pump laser (the amplitude $E_p$ and
center frequency $\omega_p$) with the Rabi frequencies
$\Omega_p=\mu_{13}E_p/\hbar$ and $k\Omega_p=\mu_{14}E_p/\hbar$,
respectively. The subband $|2\rangle$ and subbands $|3\rangle$ and
$|4\rangle$ are coupled by the Stokes laser (the amplitude $E_s$ and
center frequency $\omega_s$) with the Rabi frequencies
$\Omega_s=\mu_{23}E_s/\hbar$ and $q\Omega_s=\mu_{24}E_s/\hbar$,
respectively. The Rabi frequencies of the pump and Stokes pulses are
assumed to be Gaussian shape with the amplitude envelopes of the
form $\Omega_p = \Omega_{p0} \exp{[-(t-T_p)^2 /\tau^2]}$ and
$\Omega_s=\Omega_{s0}\exp{[-(t-T_s)^2/\tau^2]}$, respectively, where
$\tau$ is the pulse duration, and $T_{p}(T_s)$ is the time delay of
pump (Stokes) pulse. Moreover, $\Omega_{p0}$ and $\Omega_{s0}$ are
the peak values of the Rabi frequencies of pump and Stokes pulses,
respectively. For simplicity, we denote $\Omega_{p0}=\Omega_{s0}$ in
the following discussion. In addition, $k=\mu_{14}/\mu_{13}$ and
$q=\mu_{24}/\mu_{23}$ present the ratios between the intersubband
dipole moments of the relevant transitions, and $\mu_{ij}
(i,j=1-4,i\neq j) =\vec{\mu}_{ij}\cdot \vec{e}_{L}$ ($\vec{e}_{L}$
is the unit polarization vector of the corresponding laser field)
denotes the dipole moment for the transition between subbands
$|i\rangle$ and $|j\rangle$. In our structure, the dipole moments
ratios could be calculated to be $k=-0.70$ and $q=0.90$.

In the four-subband QW structure, the direct optical resonance
$|1\rangle(|2\rangle)\rightarrow|c\rangle$ ($|c\rangle$ is the
continuum subband) is much weaker than those mediate resonance paths
$|3\rangle(|4\rangle) \rightarrow |1\rangle$ and
$|3\rangle(|4\rangle) \rightarrow |2\rangle$, thus the influence of
the direct transition $|1\rangle(|2\rangle)\rightarrow|c\rangle$
could be ignored\cite{Fano}. In addition, when the electron sheet
density smaller than $10^{12}\ \rm cm^{-2}$ and with the actual
temperature at $10\ \rm K$, the electron-electron effects have very
small influence in our results. Under these assumptions, following
the standard processes\cite{Harris}, the system dynamics could be
described by the motion equations for the matrix elements in a
rotating frame:
\begin{eqnarray}
 \dot \rho_{11}&=& ik\Omega _p (\rho _{41}  - \rho _{14} )
 + i\Omega _p (\rho _{31}  - \rho _{13} ) + \gamma _{41} \rho
 _{44}\nonumber\\
 &&+ \gamma _{31} \rho _{33}  +\gamma_2 \rho_{22} + \frac{\eta}{2} (\rho _{34}  + \rho _{43} ), \label{eq:one}\\
 \dot \rho _{22}  &=& iq\Omega _s (\rho _{42}  - \rho _{24} )
 + i\Omega _s (\rho _{32}  - \rho _{23} ) + \gamma _{42} \rho
 _{44}\nonumber\\
 &&+ \gamma _{32} \rho _{33}  -\gamma_2 \rho_{22}+ \frac{\eta}{2} (\rho _{34}  + \rho _{43} ), \label{eq:two}\\
 \dot \rho _{33}  &=& i\Omega _p (\rho _{13}  - \rho _{31} )
 + i\Omega _s (\rho _{23}  - \rho _{32} ) - \gamma _3 \rho
 _{33}\nonumber\\
 &&- \frac{\eta}{2} (\rho _{34}  + \rho _{43} ) \label{eq:three}\\
 \dot \rho _{44}  &=& ik\Omega _p (\rho _{14}  - \rho _{41} )
 + iq\Omega _s (\rho _{24}  - \rho _{42} ) - \gamma _4 \rho _{44}\nonumber\\
 &&- \frac{\eta}{2} (\rho _{34}  + \rho _{43} ) \label{eq:four}\\
 \dot \rho _{12}  &=& -[i(\Delta _p  - \Delta _s )+
 \frac{\Gamma_{12}}{2}]\rho _{12}\nonumber\\
 &&+ ik\Omega _p \rho _{42}+ i\Omega _p \rho _{32}  - iq\Omega _s \rho
 _{14} - i\Omega _s \rho _{13}  \label{eq:five}\\
 \dot \rho _{13}  &=&  - (i\Delta _p  + \frac{\Gamma _{13}}{2} )\rho_{13}
 + ik\Omega _p \rho _{43}  - i\Omega _s \rho _{12}\nonumber\\
 &&+ i\Omega _p (\rho _{33} - \rho _{11} ) - \frac{\eta}{2} \rho _{14}  \label{eq:six}\\
 \dot \rho _{14}  &=&  - [i(\Delta _p  - \omega _{43} )
 + \frac{\Gamma _{14}}{2}]\rho _{14}  + i\Omega _p \rho _{34} \nonumber\\
 &&- iq\Omega _s \rho _{12}+ ik\Omega _p (\rho _{44}  - \rho _{11} ) - \frac{\eta}{2} \rho _{13}  \label{eq:seven}\\
 \dot \rho _{23}  &=&  - (i\Delta _s  + \frac{\Gamma _{23}}{2} )\rho_{23}
 - i\Omega _p \rho _{21}  + iq\Omega _s \rho _{43}\nonumber\\
 &&+ i\Omega _s (\rho _{33} - \rho _{22} ) - \frac{\eta}{2} \rho _{24}  \label{eq:eight}\\
 \dot \rho _{24}  &=&  - [i(\Delta _s  - \omega _{43} )+\frac{\Gamma _{24}}{2} ]\rho _{24}
 - ik\Omega _p \rho _{21}\nonumber\\
 &&+ i\Omega _s \rho _{34}- iq\Omega _s (\rho _{22} - \rho _{44} ) - \frac{\eta}{2} \rho _{23}  \label{eq:nine}\\
 \dot \rho _{34}  &=&  - ( - i\omega _{43}  + \frac{\Gamma _{34}}{2})\rho _{34}
 - ik\Omega _p \rho _{31}  + i\Omega _p \rho _{14}\nonumber\\
 &&- iq\Omega _s \rho _{32} + i\Omega _s \rho _{24}  - \frac{\eta}{2} (\rho _{33}  + \rho _{44} )\label{eq:ten}
\end{eqnarray}
with $\rho_{ij}=\rho_{ij}^{*}$ and the electron conservation
condition $\rho_{11}+\rho_{22}+\rho_{33}+\rho_{44}=1$. Here,
$\omega_{ij}(i,j=1-4,i\neq j)=\omega_i-\omega_j$ is the resonant
frequency between subbands $|i\rangle$ and $|j\rangle$, and
$\omega_i(i=1-4)$ is the frequency of the subband $|i\rangle$.
$\omega_{43}=\omega_4-\omega_3$ is the energy splitting between
levels $|4\rangle$ and $|3\rangle$, given by the coherent coupling
strength of the electron tunneling. $\Delta_p=\omega_p-\omega_{31}$
is the detuning of the pump laser from the resonant transition
$|1\rangle\rightarrow|3\rangle$, and $\Delta_s=\omega_s-\omega_{32}$
is the detuning between the frequency of the Stokes laser and the
transition frequency $\omega_{32}$. The population decay rates and
dephasing decay rates are added phenomenologically in the above
density-matrix. The population decay rates for subband $|i\rangle$
denoted by $\gamma_i$, are due primarily to longitudinal-optical
(LO) phonon emission events at low temperature. The total decay
rates ($\Gamma_{ij}(i\neq j)$) are given by $\Gamma_{12}=\gamma_2 +
\gamma_{12}^{dph}$, $\Gamma_{13}=\gamma_3 + \gamma_{13}^{dph}$
($\gamma_3=\gamma_{31}+\gamma_{32}$), $\Gamma_{14}=\gamma_4 +
\gamma_{14}^{dph}$ ($\gamma_4=\gamma_{41}+\gamma_{42}$),
$\Gamma_{23}=\gamma_2 + \gamma_3 + \gamma_{23}^{dph}$,
$\Gamma_{24}=\gamma_2 + \gamma_4 + \gamma_{24}^{dph}$,
$\Gamma_{34}=\gamma_3 + \gamma_4 + \gamma_{34}^{dph}$, where the
pure dephasing decay rate $\gamma_{ij}^{dph}$ is determined by
electron-electron effects, interface roughness, and phonon
scattering process. For temperature up to $10\ \rm K$ with electron
sheet densities smaller than $10^{12}\ \rm cm^{-2}$, the dephasing
rates could be estimated\cite{TIT} to be
$\gamma_{13}^{dph}=\gamma_{23}^{dph}=0.32\ \rm meV$,
$\gamma_{14}^{dph}=\gamma_{24}^{dph}=0.30\ \rm meV$,
$\gamma_{34}^{dph}=0.31\ \rm meV$ and $\gamma_{12}^{dph}=0.47\times
10^{-9}\ \rm meV$. For our QWs considered, the population decay
rates turn out to be $\gamma_3=1.58\ \rm meV$, $\gamma_4=1.50\ \rm
meV$, and $\gamma_2=2.36\times 10^{-9}\ \rm meV$.

The Fano interference factor $\eta=\sqrt{\gamma_3\gamma_4}$
represents the mutual coupling of subbands $|3\rangle$ and
$|4\rangle$ arising from the tunneling to the continuum through the
thin barrier. We define $\epsilon = \eta /\sqrt{\Gamma_{13}
\Gamma_{14}}$ for assessing the strength of the cross coupling,
where the limit values $\epsilon=0$ and $\epsilon=1$ correspond to
no interference (negligible coupling between $|3\rangle$ and
$|4\rangle$) and perfect interference (no dephasing), respectively.
From the above estimates, we obtain $\epsilon=0.83$, which could be
augmented by decreasing the temperature which generally leads to
smaller dephasing $\gamma_{ij}^{dph}$.

\section{The coherent electron population transfer}

In this section, we will investigate the achievement of perfect
coherent electron population transfer in an asymmetric semiconductor
QW system interacting with a counterintuitively ordered pulses
(i.e., the Stokes pulse precedes the pump pulse), by properly
adjusting the QW structure parameters, such as the sign of quantum
interference due to resonance tunneling, the splitting of the
excited doublet(the coupling strength of the tunneling), as well as
the coupling strength of Fano-type interference.

\begin{figure}
\includegraphics[width=7cm, height=5cm]{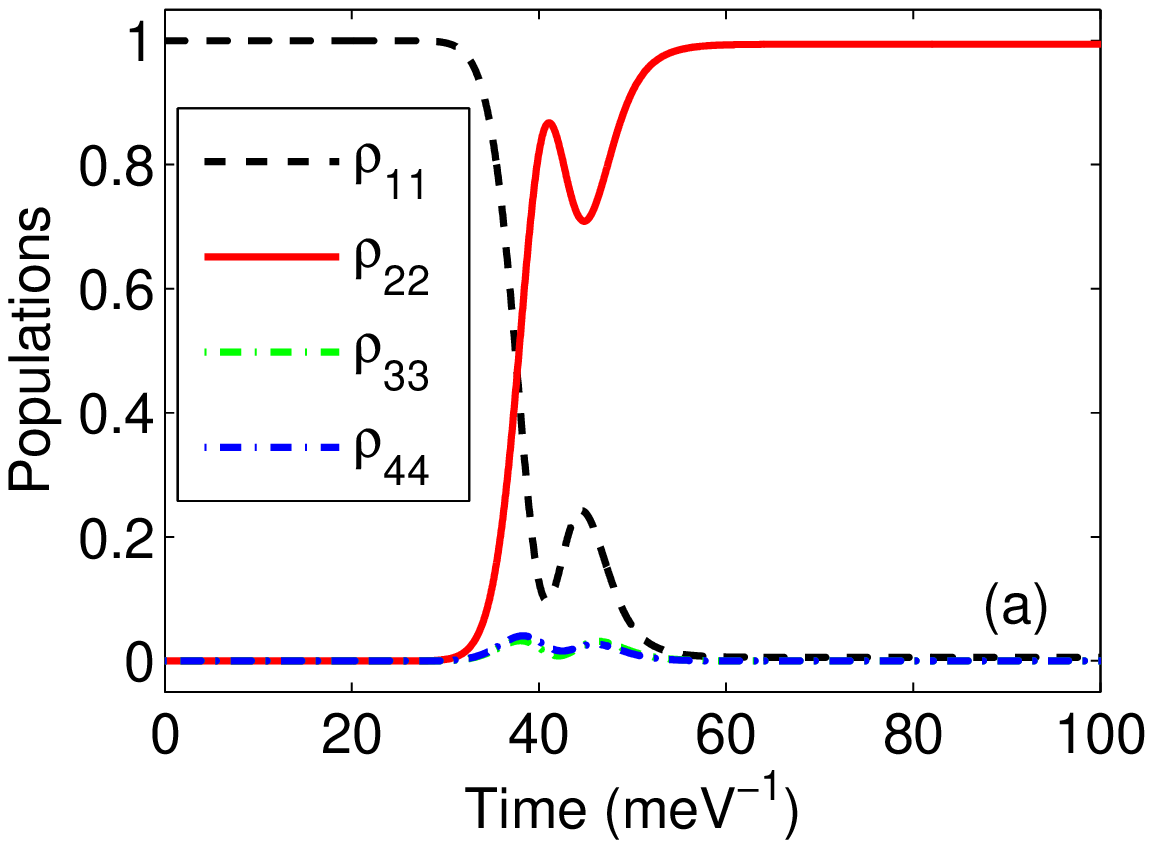}
\includegraphics[width=7cm, height=5cm]{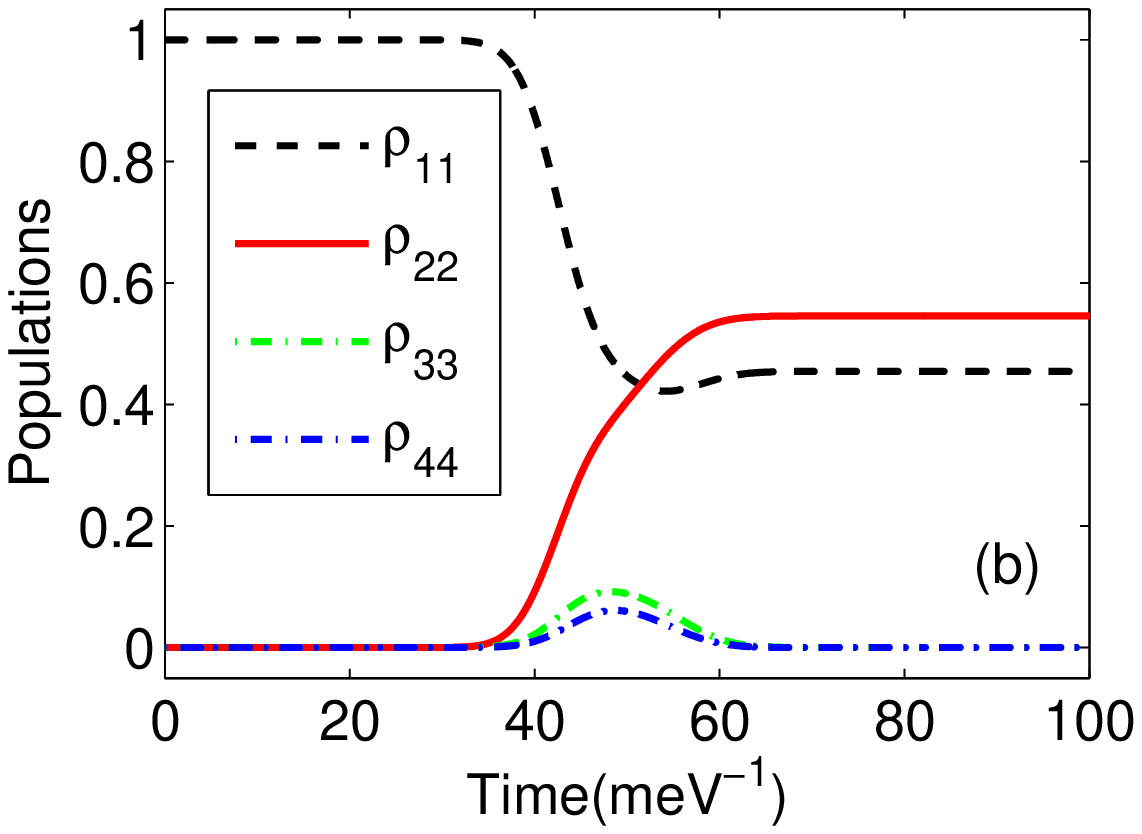}
  \caption{\label{fig:2} The time evolution of the populations in the four subbands
  $\rho_{11}$ (dash line), $\rho_{22}$ (solid line), $\rho_{33}$ (dot-dash line)
  and $\rho_{44}$ (dot-dash line) with the laser fields tuned at
  the middle point of the upper two subbands under different values
  of the dipole ratios $k=-0.70$ and $q=0.90$ in (a), and $k=0.70$ and
  $q=0.90$ in (b) with $\Omega_{p0}=\Omega_{s0}=2.6\ \rm meV$,
  $\tau=10\ \rm meV ^{-1}$, $T_s=30\ \rm meV ^{-1}$,
  $T_p=50\ \rm meV ^{-1}$, $\omega_{43}=11.76\ \rm meV$,
  $\gamma_3=1.58\ \rm meV$, $\gamma_4=1.50\ \rm  meV$ and
  $\gamma_2=2.36\times 10^{-9}\ \rm meV$.}
\end{figure}

Without the consideration of the electron dephasing and population
decays of the excited doublet, the four-subband QW structure is
similar to the excited-doublet four-level atom system \cite{Gong,
Li, Zhu}, in which some\cite{Gong} of us have demonstrate that the
dark states could exist with the contribution from the excited
doublet, and the behavior of adiabatic passage depends crucially on
the detunings between the laser frequencies and the atomic
transition frequency. When the pump and Stokes fields keep
two-photon resonance, but are not tuned at the midpoint of the
excited doublet, only one dark state exists, which is
\begin{equation}
|\phi_0\rangle=\cos{\theta}|1\rangle-\sin{\theta|2\rangle},\label{eq:dark1}
\end{equation}
where $\tan{\theta}=\Omega_p/\Omega_s$, and $\theta$ is the mixing
angle used in standard STIRAP. In the adiabatic regime, a complete
population transfer from an initial state $|1\rangle$ to another
target state $|2\rangle$ can be realized with counterintuitively
ordered pulses, just as in the three-level $\Lambda$
system\cite{CPT}. However, when both pump and Stokes fields are
tuned to the center of the two upper levels, there exist two dark
states, one is the trapped state of Eq.~(\ref{eq:dark1}), and the
other is
\begin{equation}
|\phi_1\rangle=\sin{\theta}\sin{\varphi}|1\rangle+\cos{\theta}\sin{\varphi|2\rangle}
+\frac{\cos{\varphi}}{\sqrt{2}}(|3\rangle-|4\rangle),\label{eq:dark2}
\end{equation}
where
$\tan{\varphi}=\omega_{43}/2/\sqrt{2\left[\Omega_s^2+\Omega_p^2\right]}$,
and $\varphi$ is an additional mixing angle related to the energy
separation of the excited doublet. For a counterintuitively ordered
pulses, due to the nonadiabatic coupling between the two degenerate
states $\phi_0$ and $\phi_1$, the system vector $\Psi(\infty)$ is a
mixture of the two bare states $|1\rangle$ and $|2\rangle$, as
follows
\begin{equation}
|\Psi(\infty)\rangle=\sin{\gamma_f}(\infty)|1\rangle-\cos{\gamma_f}(\infty)|2\rangle,\label{eq:dark3}
\end{equation}
where $\gamma_f=\int_{-\infty}^t(d\theta/dt')\sin{\varphi}dt'$ is
the Berry phase\cite{Gong}. Obviously, an arbitrary superposition of
atomic states can be prepared when $\omega_{43}$ is comparable to
$\Omega_p$ and $\Omega_s$. It can also be seen from
Eq.~(\ref{eq:dark3}) that, if $\omega_{43}$ is much larger than
$\Omega_p$ and $\Omega_s$, then $\varphi$ nearly equals $\pi/2$, and
almost no population transfer can occur; while if the energy
separation of the doublet $\omega_{43}$ is far smaller than the peak
Rabi frequencies $\Omega_p$ and $\Omega_s$, then $\varphi$ is nearly
equal to zero, and the population transfer behaves almost in the
same manner as that in the $\Lambda$-type three-level system.

It should be emphasized that the results obtained in
Ref.~\cite{Gong} are based on the relationship
$\frac{\Omega_p}{k\Omega_p} = \frac{\Omega_s} {q\Omega_s}$, which
means $k=q$. Indeed, the condition $k=q$ could be approximately
fulfilled, when we design the asymmetric double QWs with the
continuum adjacent to the shallow well, under which both wave
functions of subbands $|3\rangle$ and $|4\rangle$ are
symmetric\cite{Fano-sign}. However, the coherent population transfer
could be suppressed due to the destructive interference when the
dephasing and population decays of the excited doublet are taken
into account[Fig.~\ref{fig:2}(b)], as analyzed in Ref.~\cite{Zhu}
within the excited doublet four-level atomic system with spontaneous
decay-induced coherence. In our structure, we design an asymmetric
QW with the continuum adjacent to the deep well, owing to resonant
tunneling, the wave functions of subbands $|3\rangle$ and
$|4\rangle$ are symmetric and asymmetric combinations of
$|se\rangle$ and $|de\rangle$(see Fig.~\ref{fig:1}). As a result,
$kq<0$, i.e., the relationship $k=q$ can't be held. When the pump
and Stokes fields keep two-photon resonance, no dark states exist in
our structure. However, when the electron dephasing and population
decays of the excited doublet are taken into account, the almost
complete electron population transfer from an initial subband to
another target subband could be realized due to the constructive
interference from resonance tunneling with a finite pulse area.

By numerically solving the density matrix equations
[Eqs.~(\ref{eq:one})-(\ref{eq:ten})], we will investigate the
effects of the sign of resonance tunneling on the electron
population transfer in a four-subband quantum well structure with a
continuum adjacent to the deep or shallow well.
Figure.~\ref{fig:2}(a) presents the time evolution of the
populations in the four subbands of QWs [Fig.~\ref{fig:1}] with the
element ratios $k=-0.70$ and $q=0.90$, for the pump and Stokes
fields keeping the two-photon resonance and tuned midway between the
excited doublet $|3\rangle$ and $|4\rangle$. The fields parameters
are taken as $\Omega_{p0}=\Omega_{s0}=2.6\ \rm meV$, $\tau=10\ \rm
meV ^{-1}$, $T_s=30\ \rm meV ^{-1}$, and $T_p=50\ \rm meV ^{-1}$,
and the Rabi frequency is quite smaller than the energy splitting of
the excited doublet $\omega_{43}=11.76\ \rm meV$. For comparison,
Figure.~\ref{fig:2}(b) gives the time evolution of the populations
in four-subband QWs with the continuum adjacent to the shallow well,
which is similar to that in Fig.1b in Ref.~\cite{Fano-sign}, with
the element ratios $k=0.70$ and $q=0.90$. It can be seen that the
electron population transfer is dramatically modified by the sign of
the interference(destructive or constructive) due to resonant
tunneling. In the case of $k=0.70$ and $q=0.90$, in which the dipole
moments between the two upper subbands and each of the two lower
subbands are parallel, only a part of populations can be transferred
from subband $|1\rangle$ to subband $|2\rangle$
[Fig.~\ref{fig:2}(b)]. This phenomenon is due to the destructive
interference via spontaneous decay pathways which prevents the
electron population in subband $|1\rangle$ from being excited by the
pump field. That is to say, the asymmetric QW structure with the
continuum adjacent to the shallow well is not suitable for the
investigation of electron population transfer with the pump and
Stokes fields tuned midway between the excited doublet. However, in
contrast, by constructing the continuum adjacent to the deep well
through the thin barrier(see in Fig.~\ref{fig:1}), the sign of the
transition matrix element changes as $k=-0.70$, and $q=0.90$. When
the pump and Stokes fields keep the two-photon resonance and are
tuned at the center of the excited doublet, an almost complete
electron population transfer could be realized due to the
constructive interference via resonant tunneling which suppresses
the effects of the quantum coherence of spontaneous decays, as shown
in Fig.~\ref{fig:2}(a). Therefore, the perfect electron population
transfer could be achieved by suitably modulating the sign of the
interference induced by resonant tunneling, which could be obtained
by varying the direction of tunneling from the excited doublet to a
common continuum\cite{Fano-sign}.

\begin{figure}[htbp]
\includegraphics[width=7cm, height=5.5cm]{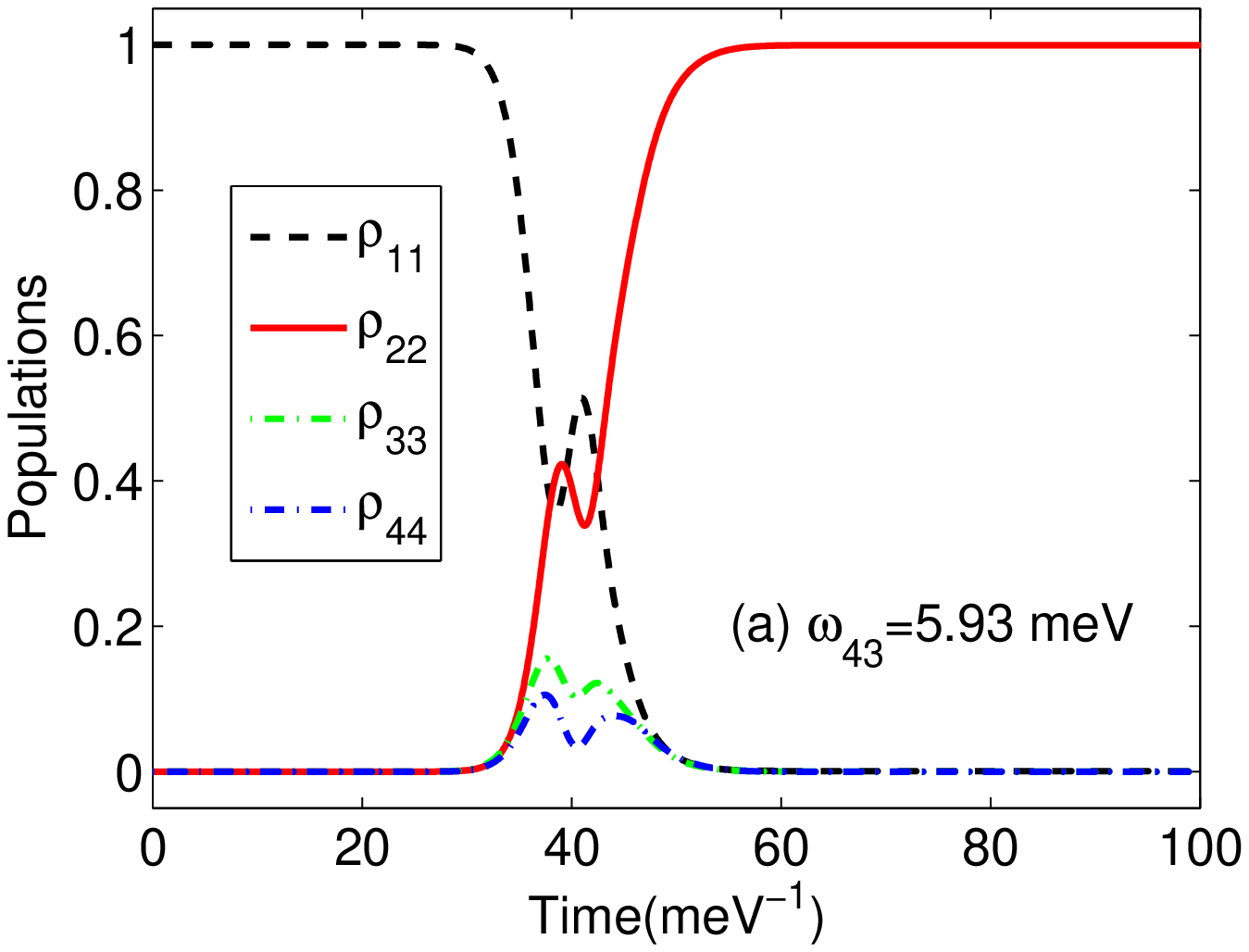}
\includegraphics[width=7cm, height=5.5cm]{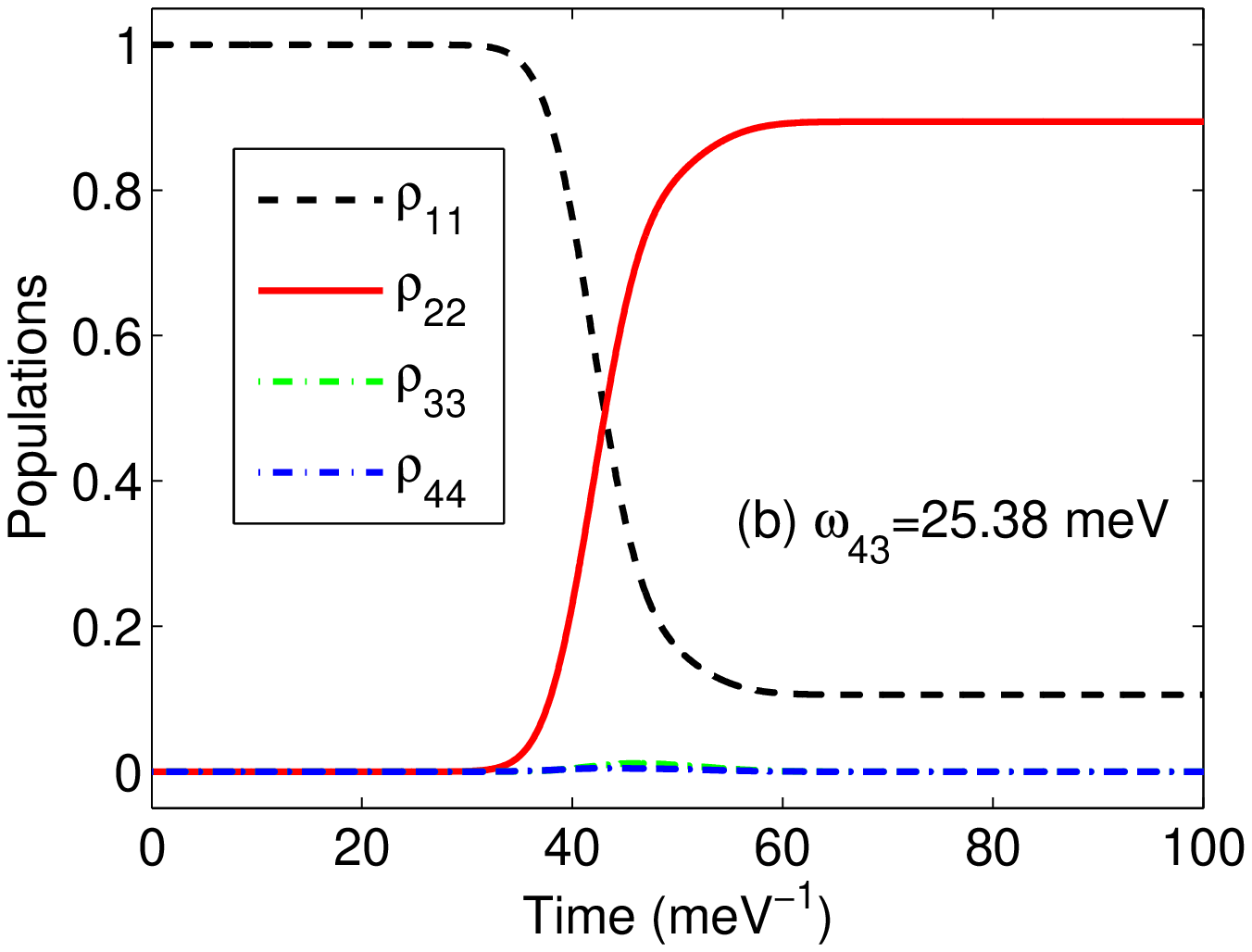}
  \caption{\label{fig:3} The time evolution of the populations
  in the four subbands $\rho_{11}$ (dash line), $\rho_{22}$ (solid line),
  $\rho_{33}$ (dot-dash line) and $\rho_{44}$ (dot-dash line) with the
  laser fields tuned at the middle point of the upper two subbands under
  different splittings of the doublet, (a) $\omega_{43}=5.93\ \rm meV$,
  $k=-0.59$ and $q=1.20$, (b) $\omega_{43}=25.38\ \rm meV$, $k=-0.61$ and
  $q=0.56$. The other parameters are the same as in Fig.~\ref{fig:2}(a).}
\end{figure}

\begin{figure}[htbp]
\includegraphics[width=7cm, height=5.5cm]{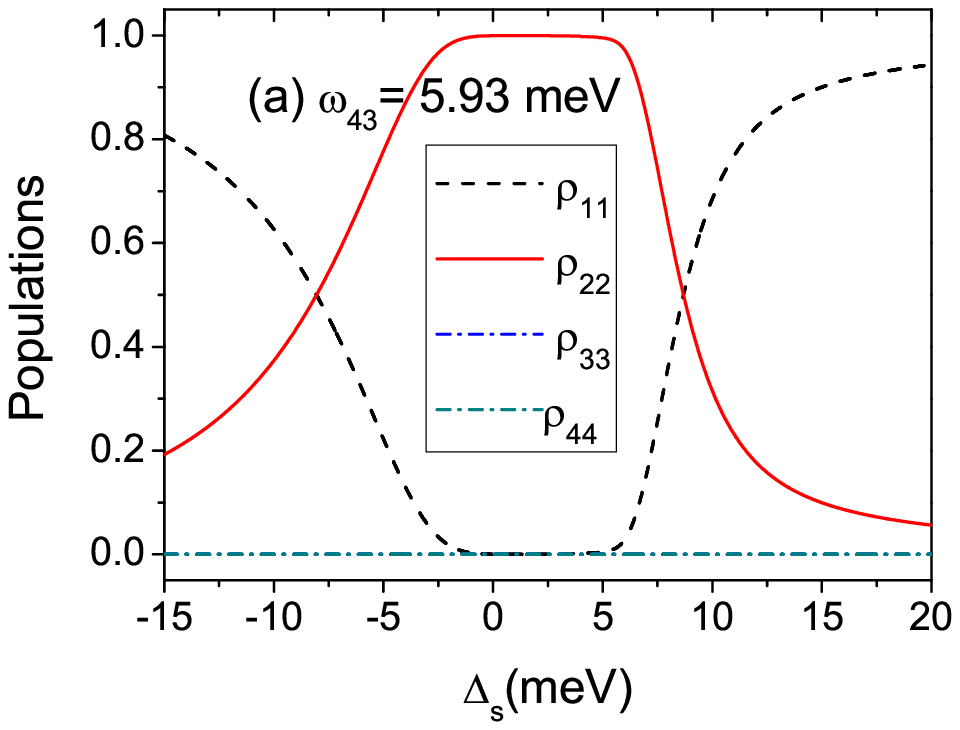}
\includegraphics[width=7cm, height=5.5cm]{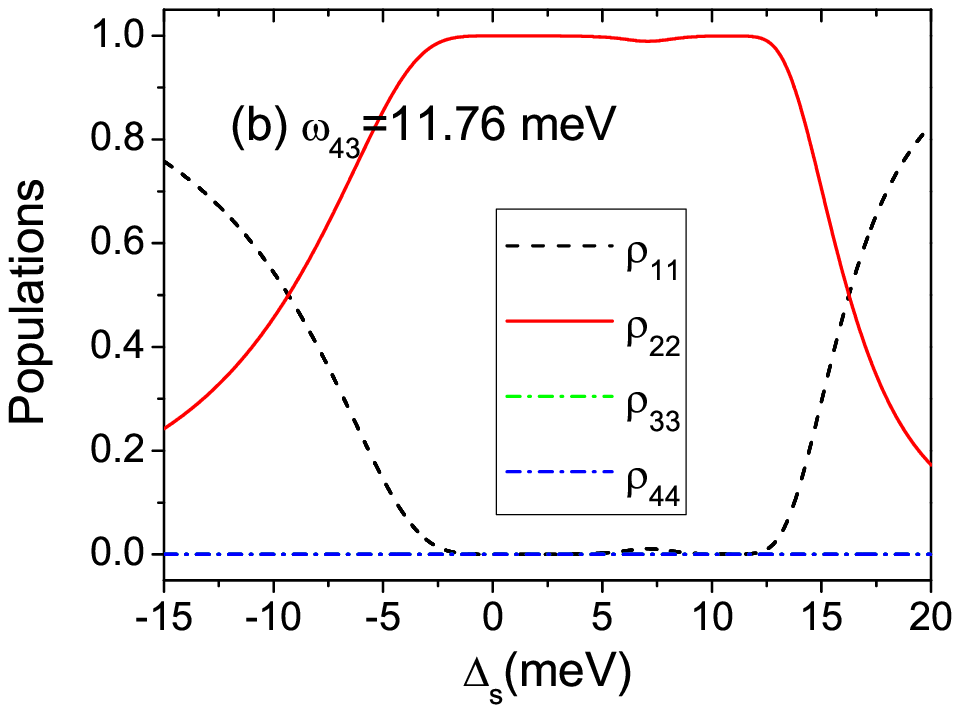}
\includegraphics[width=7cm, height=5.5cm]{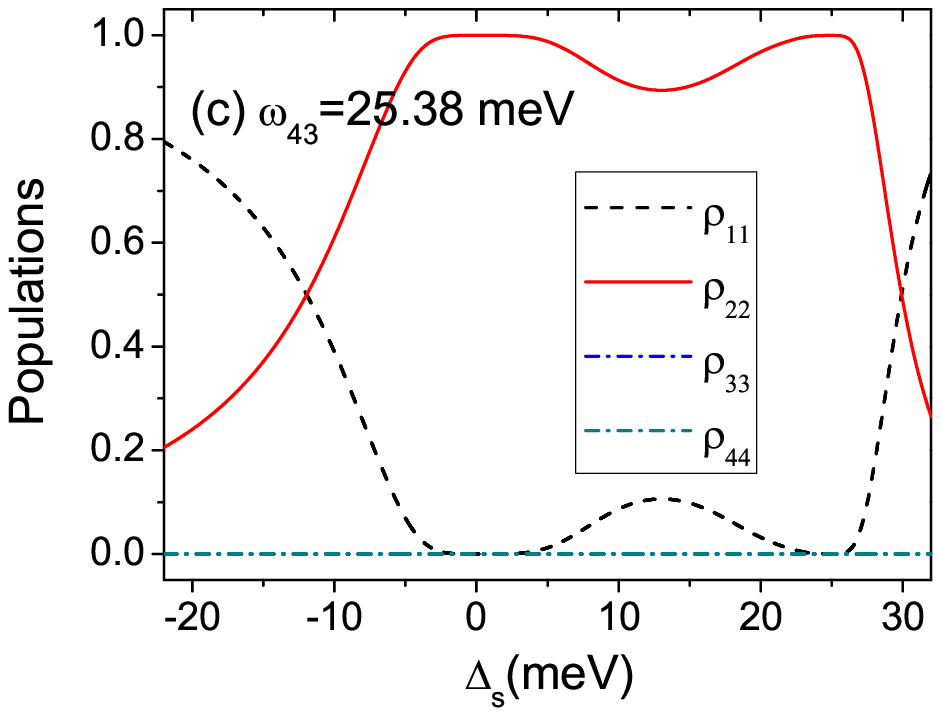}
  \caption{\label{fig:4} The final populations in the four subbands
  $\rho_{11}$ (dash line), $\rho_{22}$ (solid line), $\rho_{33}$ (dot-dash line)
  and $\rho_{44}$ (dot-dash line) as a function of the one-photon detunings of
  the pump and Stokes lasers $\Delta_{p}=\Delta_{s}$, with different splittings of the doublet,
  (a) $\omega_{43}=5.93\ \rm meV$, $k=-0.59$ and $q=1.20$, (b)$\omega_{43}=11.76\ \rm meV$,
  $k=-0.70$ and $q=0.90$, (c) $\omega_{43}=25.38\ \rm meV$, $k=-0.61$ and $q=0.56$.
  The other parameters are the same as in Fig.~\ref{fig:2}(a).}
\end{figure}

Then, we analyze how the energy splitting of the excited doublet,
which is given by the coupling strength of the tunneling, effects
the electron population transfer behavior. For our structure, the
splitting on resonance is given by the coupling strength and could
be controlled by adjusting the height and width of the tunneling
barrier. While keeping other parameters fixed except the width of
the tunneling barrier between different quantum wells, we present
the time evolution of the populations in the four subbands with
different splittings of the excited doublet in Fig.~\ref{fig:3}.
Compared with Fig.~\ref{fig:2}(a), it is can be seen that, the
perfect electron population transfer occurs when the splitting
$\omega_{43}$ is much smaller [$\omega_{43}=5.93\ \rm meV$ in
Fig.~\ref{fig:3}(a)]. As further decreasing the splitting [even
smaller than the peak Rabi frequencies $\Omega_{p}$ and
$\Omega_{s}$], the electron populations could also be completely
transferred from the initial subband $|1\rangle$ to the target
subband $|2\rangle$, which is not shown here. While with increasing
the splitting of the excited doublet, the electron population
transfer becomes lower, such as $\omega_{43}=25.38\ \rm meV$ in
Fig.~\ref{fig:3}(b). The results are in accordance with the analysis
from Eq.~(\ref{eq:dark3}), which could also be well understood from
the final populations as a function of the one-photon resonance
satisfied under different values of the splittings of excited
doublet, as shown in Fig.~\ref{fig:4}.

From Fig.~\ref{fig:4}, we find that nearly complete electron
population transfer from the initial subband $|1\rangle$ to the
target subband $|2\rangle$ at the middle point of the two one-photon
resonances with smaller splitting of excited doublet
[$\omega_{43}=5.93\ \rm meV$ in Fig.~\ref{fig:4}(a) and
$\omega_{43}=11.76\ \rm meV$ in Fig.~\ref{fig:4}(b)]. As the energy
splitting of the excited doublet are much larger than the laser
field Rabi frequency [$\omega_{43}=25.38\ \rm meV$], the electron
population transfer efficiency at the middle point decreases with
the limited pulse area. Therefore, the almost complete electron
population transfer could be controlled simply by adjusting the
energy splitting of the excited doublet, i.e., the width or the
height of the tunneling barrier. Therefore, the almost electron
population transfer to an target subband in QWs could be controlled
by appropriately adjusting the tunneling barrier.

\begin{figure}[htbp]
 \includegraphics[width=7cm, height=5.5cm]{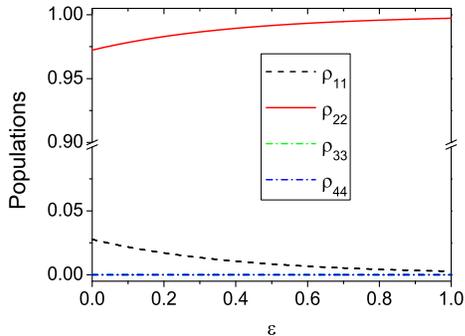}
   \caption{\label{fig:5} The final populations in the four subbands $\rho_{11}$
   (dash line), $\rho_{22}$ (solid line), $\rho_{33}$ (dot-dash line) and
   $\rho_{44}$ (dot-dash line) as a function of the Fano interference
   factor $\varepsilon$. The other parameters are the same as in Fig.~\ref{fig:2}(a).}
\end{figure}

In addition, the effects of the strength or quality of the Fano-type
interference on electron population transfer behavior could be
clearly seen from Fig.~\ref{fig:5}. From this figure, we can see
that the population transfer efficiency increases as the strength of
interference $\varepsilon$ increasing. The reason is that, the
increase of the strength of the interference, which means the
decrease of the dephasing process, enhances the coherence from the
constructive quantum interference. As a result, we can achieve an
almost complete electron population transfer via properly increasing
the Fano interference, which could be achieved by decreasing the
temperature.

It should be noted that a change in the splitting of excited doublet
may be connected to a slight change in dephasing and decay rates.
However, via properly adjusting other physical variables such as
temperatures, interface roughness scattering and so on, we believe
that some experimental scientists have adequate wisdom to keep them
unchanged. As a matter of fact, we have tried several other values
of dephasing rates $\gamma^{dph}_{12}, \gamma^{dph}_{13},
\gamma^{dph}_{14}, \gamma^{dph}_{23}, \gamma^{dph}_{24},
\gamma^{dph}_{34}$, and population decay rates $\gamma_{2},
\gamma_{3}, \gamma_{4}$ with the changes of the splitting of the
excited doublet, similar coherent population transfer behavior could
also be obtained for these different choices. In view of this, we
could keep dephasing and decay rates fixed in Figs.~\ref{fig:3} and
\ref{fig:4}.

\section{conclusions}

In conclusion, we have investigated the controllability of coherent
electron population transfer in an asymmetric double QW structure
with a common continuum interacting with counterintuitively ordered
pulses. Based on the numerical results of the motion equations of
element moments, the perfect electron population transfer could be
achieved by efficiently optimizing the semiconductor QW structure
parameters, such as the direction and the coupling strength of the
tunneling from the excited doublet to the common continuum, and the
coupling strength of the Fano interference. It is more practical to
control the population transfer behavior in the QW systems than in
the atomic systems because of its flexible design and the
controllable interference strength.

\begin{acknowledgements}
This work is supported by the National Natural Sciences Foundation
of China (Grant Nos. 60708008, 10874194, 60978013, 60921004), and
the Key Basic Research Foundation of Shanghai (Grant No.
08JC1409702).
\end{acknowledgements}

\end{document}